\documentclass[aps,prb,twocolumn,showpacs,groupedaddress]{revtex4-1}
\usepackage{latexsym}
\usepackage[english]{babel}
\usepackage{graphicx}
\usepackage{amsfonts}
\usepackage{amsmath}
\usepackage{amssymb}
\usepackage{slashed}
\usepackage{verbatim}
\usepackage{feynmf}
\usepackage{mathrsfs}
\usepackage{amsmath,amssymb}
\usepackage{epsfig}
\usepackage{relsize}
\usepackage{graphicx}
\input epsf
\usepackage{bbm}
\usepackage{color}
\usepackage{feyn}

\begin{document}

\title{Ferromagnetism in $ABC$-trilayer graphene}

\author{Richard Olsen}
\email{richard.olsen.75@gmail.com}
\author{Ralph van Gelderen}
\author{C. Morais Smith}
\affiliation{Institute for Theoretical Physics, Utrecht University, Leuvenlaan 4, 3584 CE Utrecht, The Netherlands}

\date{\today}

\pacs{75.70.Cn, 73.22.Pr, 73.20.At}

\begin{abstract}
In this article we study the ferromagnetic behavior of $ABC$-stacked trilayer graphene. This is done using a nearest-neighbor tight-binding model, in the presence of long-range Coulomb interactions. For a given electron-electron interaction $g$ and doping level $n$, we determine whether the total energy is minimized for a paramagnetic or ferromagnetic configuration of our variational parameters. The $g$ versus $n$ phase diagram is first calculated for the unscreened case. We then include the effects of screening using a simplified expression for the fermion bubble diagram. We show that ferromagnetism in $ABC$-trilayer graphene is more robust than in monolayer, in bilayer, and in $ABA$-trilayer graphene. Although the screening reduces the ferromagnetic regime in $ABC$-trilayer graphene, the critical doping level remains one order of magnitude larger than in unscreened bilayer graphene.
\end{abstract}

\maketitle

\section{Introduction}

Within a decade after the discovery of graphene flakes by mechanical exfoliation,\cite{Novo04} numerous methods have been developed to create larger and cleaner samples, realized both as single layers and as stacked layers of graphene.\cite{SooKim09,Cao10,Bae10,Ted09,Gom12,Reina09}

Early on, it was realized that stacked graphene layers behave differently than both a single layer and 3D graphite. For example, in bilayer graphene the dispersion is quadratic instead of linear and the electrons behave as massive chiral particles, which is a completely new type of particle. Few-layer graphene is still a 2D system, hence the quantum Hall effect can be observed. For monolayer graphene, the plateaus in the Hall conductivity are located at half integer multiples of $4 e^2/h$,\cite{Novo05} originating from a Landau level at zero energy which is half filled by electrons and half filled by holes. In bilayer graphene, this particular Landau level has an extra degeneracy resulting in Hall plateaus at integer values of $4 e^2/h$ and a quantum Hall effect that is different from the one in a monolayer as well as from the quantum Hall effect found in usual two dimensional electron gases.\cite{McCann06} In addition to the number of layers, the order of the stacking also influences the physical properties significantly.

In multilayer graphene, the different layers can have three distinct orientations with respect to the bottom one. Bernal stacking (or $AB$ stacking) is the configuration in which the $\mathcal{B}$ sublattice of the odd layers are opposite to the $\mathcal{A}$ sublattice of the even layers. The Hamiltonian of a system with an even number $2N$ of layers can be rewritten in a block diagonal form, where the $N$ different blocks are bilayer-like Hamiltonians. The blocks can be linked by hopping parameters that couple lattice sites on next-nearest planes. For an odd number ($2N+1$) of layers, one of the blocks is the monolayer Hamiltonian. Therefore, these systems have a linear band in addition to the $N$ parabolic ones.\cite{McCa10a}

\begin{figure}[t]
\includegraphics[width=.28\textwidth]{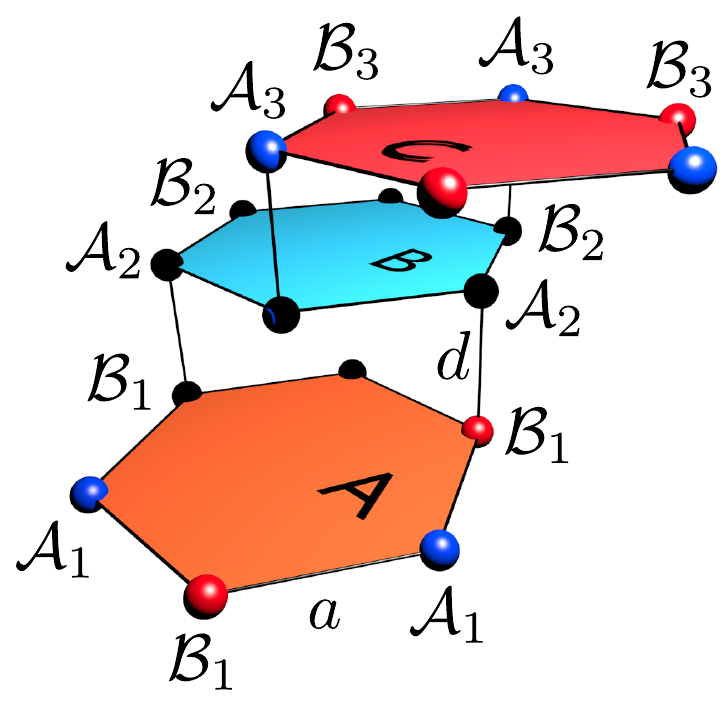}
\caption{(Color online) Atomic structure of $ABC$-trilayer graphene.}
\label{fig1}
\end{figure}

In $ABC$ stacked multilayer graphene, the $\mathcal{B}$ sublattice of each layer lies opposite to the $\mathcal{A}$ sublattice of the layer above it, but opposite to the honeycomb centers in the layer beneath it (see Fig.~\ref{fig1}). Since electrons that are placed oppositely in two bordering planes dimerize, resulting in an energy shift away from zero, these multilayers can, for low energies, be described by a $2\times 2$ effective matrix Hamiltonian, which is governed by the indirect (effective) hopping between the two atoms in the outer planes that have no neighbor in the adjacent layer. This effective hopping is a process consisting of $N-1$ interplane nearest-neighbor hoppings, combined with $N$ in-plane nearest-neighbor hoppings, resulting in an energy dispersion around the $K$-points, $E_N\sim v_F^N k^N /t_\perp^{(N-1)}$.\cite{McDo08a}

A tight-binding approach for an increasing number of layers should in principle include hopping between more distant carbon atoms. The long known Slonczewski-Weiss-McClure (SWMc) model\cite{SW58,McC57} accounts for next-nearest-neighbor hopping, as well as hopping between next-nearest planes. In fact, trilayer graphene can be used to obtain the values of the different hopping parameters by fitting experimental data to the SWMc model.\cite{TayHer11} However, often it is sufficient to take into account only the intra- and interplane nearest-neighbor hopping parameters.

Recent experimental and theoretical studies of trilayer graphene have shown that magnetotransport and electronic transport properties,\cite{Jhang11} thermoelectric transport properties\cite{Ma12}, and chiral tunneling\cite{Kumar12} indeed depend on the stacking order. Furthermore, one can open a sizeable bandgap in $ABC$-stacked trilayers ($120$ meV) by applying an external electric field, while for an $ABA$-trilayer no gap is observed under the same conditions.\cite{Hung11}

Extensive research into the band structure of $ABC$-multilayer graphene has been done recently using an effective mass approximation.\cite{Koshino2} It was found that the electron and hole bands touching at zero energy support chiral quasiparticles characterized by a Berry phase of $N\pi$ for $N$ layers. The phonon spectrum of $ABC$-stacked graphene has been investigated theoretically using density functional theory\cite{Yan08} and experimentally by using infrared absorption spectroscopy, where the intensities have been found to be much stronger than that of bilayer graphene.\cite{Li2012} Using magnetic fields up to 60T, there has been evidence of the integer quantum Hall effect in trilayer graphene.\cite{Kumar11} The Hall resistivity plateaus have been reproduced by using a self-consistent Hartree calculation on $ABC$-stacked graphene.\cite{Kumar11} It has been suggested that the differences in the quantum Hall effect between $ABC$- and $ABA$- stacking might be used to identify the stacking order of high-quality trilayer samples.\cite{Zhang2012} By using infrared absorption spectroscopy, it has been shown that the optical conductivity spectra for $ABC$- and $ABA$- stacked graphene differs considerably.\cite{Mak2010} These optical properties have been calculated and reproduced in the framework of a tight-binding model.\cite{Yan2011} Finally, it can be mentioned that high-resolution transmission microscopy of $ABC$-stacked trilayer graphene on a SiC surface has successfully provided information on the interlayer distances of $ABC$-trilayer graphene.\cite{Borysiuk2011}

In this article we investigate the magnetic properties of $ABC$-trilayer graphene by using a nearest-neighbor tight-binding model, in the presence of long-range Coulomb interactions. For interacting electrons, the system can gain energy by aligning the spins of the electrons. This \textit{exchange mechanism} is accompanied by a cost in kinetic energy due to the Pauli principle. After fixing the doping level and interaction strength, one can use a variational approach to determine whether the system spontaneously magnetizes or remains paramagnetic. For monolayer graphene, the system only magnetizes if the interaction strength is tuned to unphysically high values. Depending on the doping level $n$, this phase transition can be first or second order.\cite{PeCaNe05a} For bilayer graphene the system can be ferromagnetic for the estimated value of the Coulomb interaction ($g=2.1$), but the electron density has to be as low as $n \sim 10^9$ cm$^{-2}$ for the material to become magnetic.\cite{NiCaNe06a} This is on the brink of what is experimentally achievable, since it is not possible to create perfectly undoped graphene in experiment, due to the formation of electron hole puddles\cite{Mart08} and impurities trapped in the substrate. In $ABA$-trilayer, the interplay between the linear and the parabolic bands opens up possibilities for both spin,- and band-ferromagnetism, but only at low electron doping.\cite{Rvg11}

\begin{figure}[t]
\includegraphics[width=.48\textwidth]{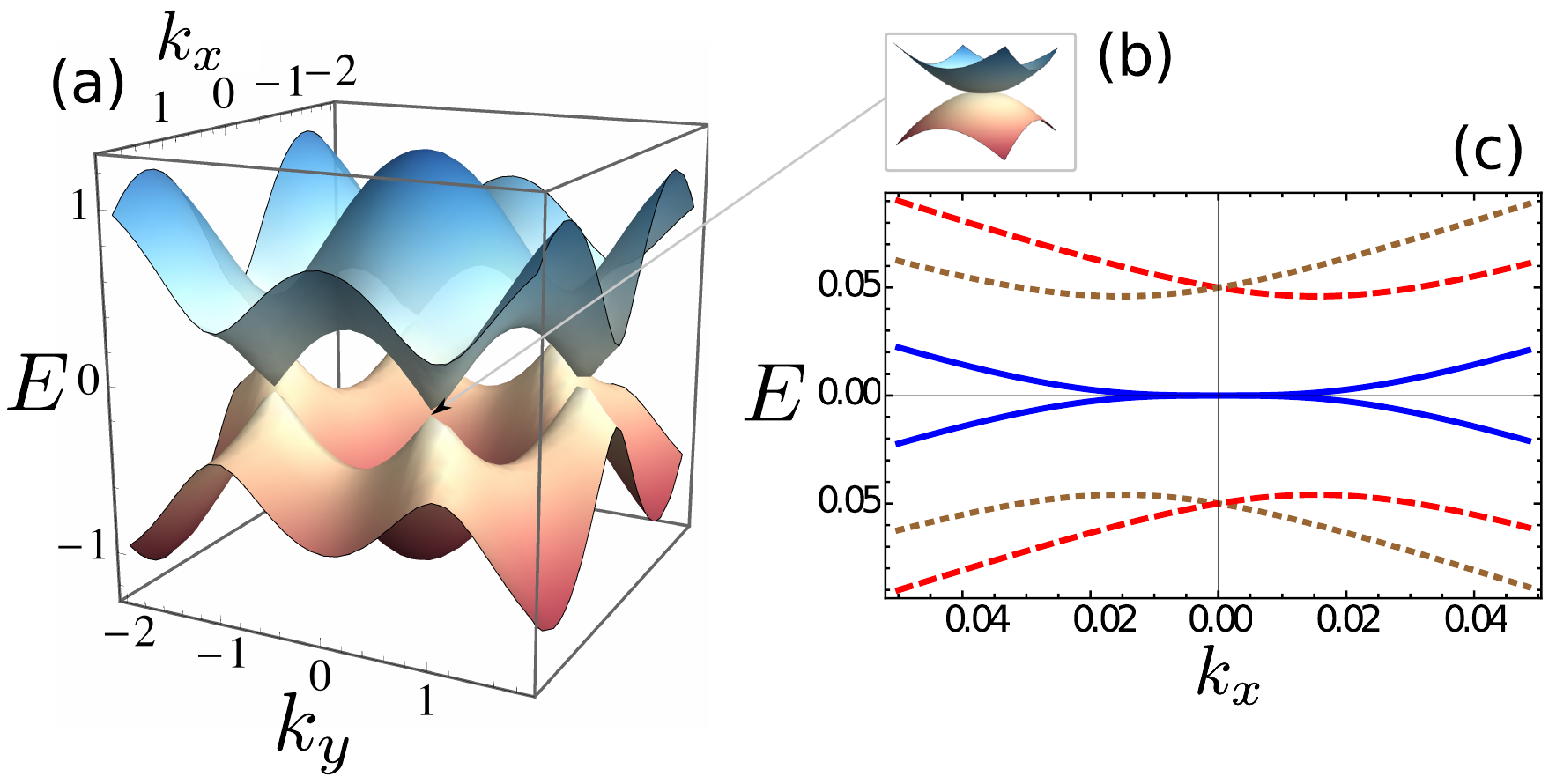}
\caption{(Color online) (a) Numerically calculated, full dispersion of $ABC$-trilayer graphene (two lowest bands). (b) Zoom-in on one of the K-points. (c) Numerically calculated low energy approximation of $ABC$-trilayer graphene dispersion (expansion around the K-point).}
\label{fig2}
\end{figure}

Although in a low energy approximation $ABC$-trilayer graphene seems to be -- in a way -- the three layer generalization of the Bernal stacked bilayer,\cite{McDo08a} it is worth a further investigation because its cubic energy dispersion is expected to enhance significantly the phase-space where the ferromagnetic regime occurs. In addition, screening should play an important role, due to the diverging density of states. Here we show that this is indeed the case: although the screening reduces the regime of parameters for the occurrence of ferromagnetism, the latter remains at least one order of magnitude more robust than in  unscreened bilayer graphene. The outline of our paper is the following: we set up the model in Sec.~\ref{sec:themodel}, present our results of the unscreened case in Sec.~\ref{sec:unscreened}, and look at the effects of screening in Sec.~\ref{sec:effectsscreening}. Our conclusions are drawn in Sec.~\ref{sec:conclusions}.

\section{The model}
\label{sec:themodel}

\begin{figure*}[thb]
\includegraphics[width=.98\textwidth]{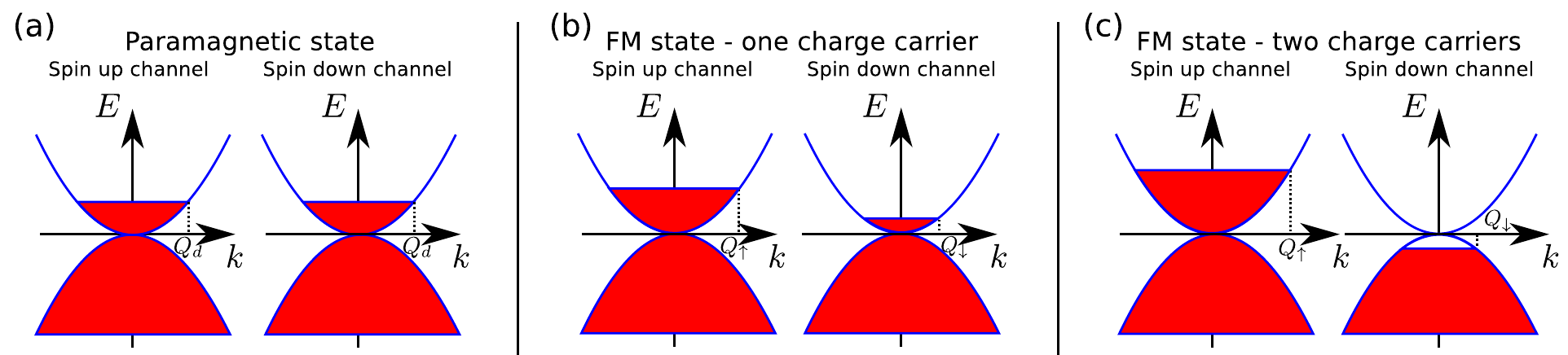}
\caption{(Color online) Sketch of the electron- (hole-) pockets for three configurations of the system - (a) paramagnetic, (b) ferromagnetic with one type of charge carrier and (c) ferromagnetic with two types of carriers.}
\label{fig3}
\end{figure*}

We use a tight-binding model which takes into account the hopping of electrons to nearest-neighbor inplane and interplane sites. In real space, the Hamiltonian is given by
\begin{equation}
	H=H_0+H_I,
\end{equation}
with the non-interacting part being
\begin{align}
	H_0 &= -t \sum_{\langle i,j\rangle,\sigma} \sum_{n=1}^3 \left[ {a}_{i,\sigma,n}^\dagger {b}_{j,\sigma,n} + \text{h.c.} \right] \\
	&\phantom{=} - t_\perp \sum_{i,\sigma} \left[ {b}_{i,\sigma,1}^\dagger {a}_{i,\sigma,2} + \text{h.c.} \right] \nonumber \\
	&\phantom{=} - t_\perp \sum_{i,\sigma} \left[ {b}_{i,\sigma,2}^\dagger {a}_{i,\sigma,3} + \text{h.c.} \right], \nonumber
\end{align}
where $i$ and $j$ label the lattice sites, $\sigma \in \{ \uparrow,\downarrow \}$ labels spin, $n \in \{1,2,3\}$ labels the layer, $t\approx 3$ eV denotes the intra-layer nearest-neighbor hopping parameter, $t_\perp \approx 0.35$ eV denotes the interlayer nearest-neighbor hopping, and the operator $c^\dagger$ ($c$) creates (annihilates) an electron on sublattice $\mathcal{C}\in\{\mathcal{A},\mathcal{B}\}$. $H_I$ is the interaction Hamiltonian. Since the stacking considered is $ABC$, the $\mathcal{A}$ sublattice in the bottom layer (layer 1) and the $\mathcal{B}$ sublattice in the top layer (layer 3) do not have direct neighbors in an adjacent layer. The electrons interact via a Coulomb interaction, which can be included in our model by the term
\begin{align}
\nonumber H_I &= \frac{1}{2} \int d^2 \mathbf{x} \, d^2 \mathbf{y} \big\{ V^\textrm{D}(\mathbf{x}-\mathbf{y}) [ \rho_1(\mathbf{x}) \rho_1(\mathbf{y})+\rho_2(\mathbf{x}) \rho_2(\mathbf{y}) \\
\nonumber &\phantom{=} +\rho_3(\mathbf{x}) \rho_3(\mathbf{y})] +V^{\textrm{ND}}(\mathbf{x}-\mathbf{y})[ \rho_1(\mathbf{x}) \rho_2(\mathbf{y}) \\
\nonumber &\phantom{=}+ \rho_2(\mathbf{x}) \rho_1(\mathbf{y})+ \rho_2(\mathbf{x}) \rho_3(\mathbf{y})+ \rho_3(\mathbf{x}) \rho_2(\mathbf{y})] \\
\label{intrealsp} &\phantom{=}+V^{\textrm{2ND}}(\mathbf{x}-\mathbf{y}) [ \rho_1(\mathbf{x}) \rho_3(\mathbf{y})+\rho_3(\mathbf{x}) \rho_1(\mathbf{y})] \big\},
\end{align}
where the density of electrons in the $n$-th layer is given by $\rho_{n}(\mathbf{x})=\sum_\sigma {\Psi}^\dagger_{\sigma,n}(\mathbf{x}){\Psi}_{\sigma,n}(\mathbf{x})$, with ${\Psi}_{\sigma,n}(\mathbf{x})\equiv (a_{\sigma,n}(\mathbf{x}),b_{\sigma,n}(\mathbf{x}))$, where $a_{\sigma,n}(\mathbf{x})$ and $b_{\sigma,n}(\mathbf{x})$ are the field operators corresponding to $a_{i,\sigma,n}$ and $b_{i,\sigma,n}$, respectively. The interaction potentials for the in-plane (D), the nearest-neighbor planes (ND) and the next-nearest-neighbor planes (2ND) are given by
\begin{align}
\nonumber V^{\textrm{D}}(\mathbf{x}-\mathbf{y})&= \frac{2\pi e^2}{\epsilon |\mathbf{x}-\mathbf{y}|}, \\
\nonumber V^{\textrm{ND}}(\mathbf{x}-\mathbf{y})&= \frac{2\pi e^2}{\epsilon \sqrt{d^2+|\mathbf{x}-\mathbf{y}|^2}}, \\
\nonumber V^{\textrm{2ND}}(\mathbf{x}-\mathbf{y})&= \frac{2\pi e^2}{\epsilon \sqrt{4 d^2+|\mathbf{x}-\mathbf{y}|^2}}.
\end{align}
In these interaction potentials, $d \approx 3.2$~\AA~is the interlayer distance, $e$ the electron charge, and $\epsilon$ the dielectric constant of the substrate.

\subsection{Kinetic energy}

After Fourier transforming and expanding the momenta around the $K$-point, the non-interacting Hamiltonian acquires the form
\begin{align}
\label{hamdens} H_0 &= \sum_{\sigma} \int d \mathbf{k} \Psi_\sigma^\dagger(\mathbf{k}) \mathcal{H}(\mathbf{k}) \Psi_\sigma ( \mathbf{k} ), \\
\nonumber  \Psi_\sigma^\dagger(\mathbf{k})&= ( a^\dagger_{\mathbf{k},\sigma,1}, b^\dagger_{\mathbf{k},\sigma,1}, a^\dagger_{\mathbf{k},\sigma,2}, b^\dagger_{\mathbf{k},\sigma,2}, a^\dagger_{\mathbf{k},\sigma,3},b^\dagger_{\mathbf{k},\sigma,3} ),
\end{align}
where $c^\dagger_{\mathbf{k},\sigma,n}$ creates a particle with momentum $\mathbf{k}$ on sublattice $c \in \{a,b\} $ in layer $n$ with spin $\sigma$, and $\mathcal{H}$ is a $6 \times 6$ matrix given by
\begin{align}
\label{matrixhamiltonian} \mathcal{H} &= \hbar v_F \left( \begin{array}{cccccc} 0 & u & 0 & 0 &0 &0 \\ u^* & 0 & -\gamma_1 &0&0&0 \\ 0&-\gamma_1&0&u&0&0 \\ 0&0&u^*&0&-\gamma_1&0 \\ 0&0&0&-\gamma_1&0&u \\ 0&0&0&0&u^*&0 \end{array} \right),
\end{align}
where $u\equiv k e^{i \phi( \mathbf{k} )}$. In the above expression, $k=\vert \mathbf{k} \vert$ is the norm of the two-dimensional momentum vector, $\phi(\mathbf{k})=\arctan\left( k_y/k_x \right)$ is the angle of the momentum vector, $\hbar v_F=(3/2) a t$ is the Fermi velocity in terms of the lattice constant $a=1.42$~\AA~and intralayer hopping parameter $t$, and $\gamma_1\equiv t_\perp/(\hbar v_F)$.

Although it is possible to write an analytic expression for the low energy approximation of the single-particle dispersion for $ABC$-trilayer graphene,\cite{Kosh10} this is not the case for the required diagonalization matrix for $\mathcal{H}$. For this reason, we calculate both numerically.  The full dispersion is shown in Fig.~\ref{fig2}(a) and (b), together with an expansion of the energy bands around the K-point (i.e. eigenvalues of Eq.~(\ref{matrixhamiltonian})), which are indeed cubic for small momenta (at small momenta $E(k) \approx \pm (v_F^3/t_\perp^2)k^3$ for the two lowest bands), see Fig.~\ref{fig2}(c).

When the system undergoes a phase transition into a ferromagnetic state, pockets of one spin configuration -- let us say up -- will be larger than the pocket of spin-down electrons [see Fig.~\ref{fig3}(a)-(b)] Moreover, it is also possible to have two types of charge carriers in the system, i.e. the formation of spin-up electron-pockets and spin-down hole pockets [see Fig.~\ref{fig3}(c)].

To compute the energy of an electron or hole pocket of size $Q_\sigma$ (see Fig.~\ref{fig3}), we have to compute the integral 
$$ \Delta K = \int_0^{E(Q_\sigma)} E \mathcal{D}(E) dE,$$
where $\mathcal{D}(E)$ is the density of states 
$$\mathcal{D}(E)=\frac{\partial N}{\partial E}= \frac{A}{4 \pi} \frac{\partial}{\partial E}\left[ k(E)^2 \right],$$ 
with $A$ denoting the area of the unit cell and $N$ is the number of states below $E$. We compute the inverse of the dispersion relation $E(k)$ numerically. Note that for small pocket sizes, $\Delta K\sim Q_\sigma^5$. When compared with monolayer graphene ($\Delta K^{m.l.}\sim Q_\sigma^3$)\cite{PeCaNe05a} and bilayer graphene ($\Delta K^{b.l.}\sim Q_\sigma^4$)\cite{NiCaNe06a} it is evident that the kinetic energy cost of an electron (hole) pocket is smaller in $ABC$-trilayer graphene than in the fewer-layered carbon structures.

\subsection{Exchange energy}

When calculating the energy contribution coming from $H_I$, the direct contribution (i.e. the Hartree term) cancels due to the positive Jellium background. The only term left is the exchange contribution (i.e. the Fock term), which favors spin alignment. However, spin alignment will result in a cost in kinetic energy due to the Pauli exclusion principle. Thus, ferromagnetism will occur or not, depending on the competition between the kinetic energy and the exchange energy.

In the Appendix, it is shown that the exchange energy of a configuration as in Fig.~\ref{fig3}, where the spin-up and the spin-down bands fill up differently, can be written in a way similar to the one in bilayer graphene,\cite{NiCaNe06a}
\begin{align}
\label{exchenergy} \frac{E_\textrm{ex}}{A} &= -\frac{1}{2} \int \frac{d\, \mathbf{k}}{(2 \pi)^2} \frac{d\, \mathbf{k}'}{(2 \pi)^2} \sum_{\sigma,a} \sum_{s=1}^6 \sum_{\alpha,\beta =1}^6 \\
\nonumber & \bigg[ \chi^s_{\alpha\beta}(\mathbf{k}',\mathbf{k}) \chi^s_{\beta\alpha}(\mathbf{k},\mathbf{k}') V_s(\mathbf{k}'-\mathbf{k}) n_{\sigma,\alpha,a}(\mathbf{k}') n_{\sigma,\beta,a}(\mathbf{k}) \bigg].
\end{align}
Here, $\alpha$ and $\beta$ label the band index and $a$  labels the valley, but we will neglect intervalley scattering and only focus on the $K$ point. $n_{\sigma,\alpha,a}(\mathbf{k})$ are the Fermi functions and the expressions for $ V_s(\mathbf{k}'-\mathbf{k})$ are given in the Appendix. In comparison with the bilayer, there are six $\chi$ matrices instead of two and they are no longer $4 \times 4$, but $6 \times 6$. Moreover, they can only be computed numerically (see the Appendix for more details).

Since we have expanded around the $K$ point, we introduce a cutoff $\Lambda=\sqrt{2 \pi/A}$ in such a way that the number of states in the Brillouin zone is conserved. Using the cutoff, we can measure momenta (and hence pocket sizes) in units of $\Lambda$ and energies in units of $\hbar v_F \Lambda (\approx 7.2$ eV$)$. This makes all our variables and parameters dimensionless and after setting $\hbar=1$, $v_F=1$, and $\Lambda=1$ they have the following values: $t=0.42$, $t_\perp=0.05$, $a=1.56$, and $d=3.7$.\cite{footnote2}

\section{Unscreened case}
\label{sec:unscreened}

\subsection{Numerical solution}

\begin{figure}[t]
\includegraphics[width=.48\textwidth]{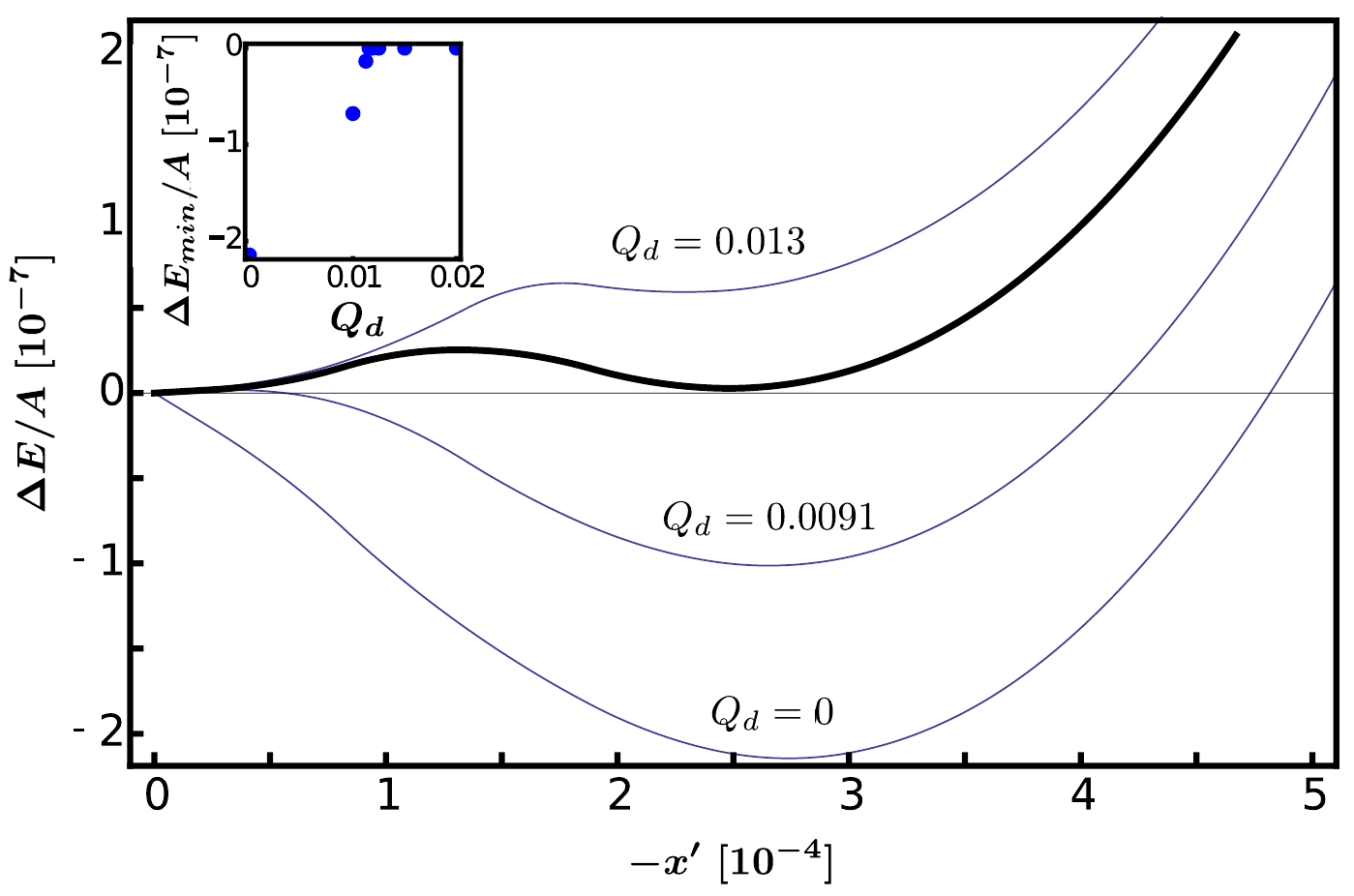}
\caption{(Color online) $\Delta E$ versus $-x'$ at electron-electron coupling $g=6$. The phase transition occurs at the doping $Q_d\approx 0.0116$ that produces the thick black curve. Inset: Minimum energy versus doping $Q_d$. The phase transition is identified by the value of $Q_d$ where $\Delta E_{min}$ first becomes non-zero.}
\label{fig4}
\end{figure}

The exchange energy $E_{ex}/A$ given by Eq.~(\ref{exchenergy}) is solved numerically using the double exponential (DE) algorithm \cite{Masatake01} (the DE algorithm is originally intended for 1D integrals, but is extended to 3D to perform the exchange integrals). Due to the singular behavior of the Coulomb potentials, the integral must undergo a series of transformations. Firstly, the integral is transformed to polar coordinates, where we introduce a cutoff $\Lambda$ for integrals over the norm of the the momentum. A change of variables is then applied, such that these integrations range from zero to one. This permits the singular behavior along $\mathbf{k}=\mathbf{k}'$ to be rotated by a Duffy coordinate transformation \cite{Duffy84}
\begin{align}
	\int_0^{2\pi} &d\theta \int_0^1 dk \int_0^1 dk' \frac{F(k,k',\theta)}{\sqrt{(k'^2-2kk'Q\cos\theta+Q^2k^2}} \nonumber \\
	&= \int_0^{2\pi} d\theta \int_0^1 dk \int_0^1 dk'' \bigg[\frac{F(k,kk'',\theta)}{\sqrt{{k''}^2-2k''Q\cos\theta+Q^2}} \nonumber \\
	&+ \frac{F(kk'',k,\theta)}{\sqrt{1-2k''Q\cos\theta+Q^2{k''}^2}} \bigg]. \label{duffytransform}
\end{align}
This formula is derived by splitting the $k'$ integration into two separate integrations from 0 to $k$ and from $k$ to 1. Making the change of variables $k'=kk''$ on the first integral leads to the first term on the right hand side of Eq.~(\ref{duffytransform}). In the second integral, with integration boundaries $k$ and $1$, the identity $\int_0^1dk\int_k^1dk'f(k,k') = \int_0^1dk\int_0^kdk'f(k',k)$ is applied. Thus, a change of variables $k'=kk''$ leads to the second term on the right hand side of Eq.~(\ref{duffytransform}).

The singularities are now confined to lines parallel to the $k$-axis. However, there are now two such lines of singularities in the integrand, located at $k''=h_1\neq 1$ and $k''=h_2\neq 1$. The lines of singularities located at $h_1$ and $h_2$ must be moved to $k''=1$ by a change of variables. After the change of variables, the integration boundaries are no longer confined to zero and one. Since the DE algorithm is only capable of handling singularities at the integration boundaries, all integrals are split at $k''=1$ (where the singularities are now located), before being performed.

The Hamiltonian matrix $\mathcal{H}$ of Eq.~(\ref{matrixhamiltonian}) is diagonalized numerically using the Jacobi diagonalization algorithm, which is extended to handle a Hermitian $6\times 6$ matrix by solving the corresponding $12 \times 12$ real symmetric matrix.\cite{NumRecip} The resulting diagonalization matrix $\mathcal{M}(\mathbf{k})$ is used inside the $\chi$ matrices of Eq.~(\ref{exchenergy}) to calculate the exchange energy, while the resulting dispersion $E(k)$ is used to calculate the kinetic energy (see Appendix for details).

The numerical diagonalization process does not provide $E^{-1}(k)$, which is needed to calculate the kinetic energy. Thus, the inverse is approximated by linear interpolation of the dispersion. Integration by parts yields
$$\Delta K = E(Q) N(E(Q)) - \int_0^{E(Q)}N(E)dE,$$
which is used in order to avoid explicit numerical evaluation of $\partial N/\partial E$.

Consider a paramagnetic state with doping $Q_d$ and a ferromagnetic state with electron (or hole) pockets $Q_{\uparrow}$ and $Q_{\downarrow}$. Then, the kinetic energy difference is calculated by
$$\frac{\Delta E_{kin}}{A}=\frac{1}{A} \left[\Delta K(Q_{\uparrow}) + \Delta K(Q_\downarrow) - 2\Delta K(Q_d)\right].$$
The difference in exchange energy $\Delta E_{ex}/A$ is calculated by subtracting $E_{ex}/A$ of the paramagnetic state from the corresponding energy of the ferromagnetic state. For an unperturbed system, both spin channels are filled up to the Fermi-momentum $Q_d$ [see Fig.~\ref{fig3}(a)]. Due to the exchange mechanism, the system can prefer a ferromagnetic state with either one type of carrier or two types of carriers [see Fig.~\ref{fig3}(b)-(c)]. These perturbations are parameterized by the variable $x$, which is positive for one type of carrier and given by
$$Q_{\uparrow}^2=2 Q_d^2-x, \qquad Q_{\downarrow}^2=x.$$
For two types of carriers, $x$ is defined to be negative and parameterizes the electron and hole pocket as
$$Q_{\uparrow}^2=2 Q_d^2+|x|, \qquad Q_{\downarrow}^2=|x|,$$
where we assume the electron pocket in the spin-up channel. Using this parametrization, particle conservation is satisfied. It is convenient to introduce $x'\equiv x-Q_d^2$, such that $x'=0$ represents the unperturbed state (i.e. $Q_\uparrow=Q_\downarrow=Q_d$). Then, $\Delta E / A = \Delta E_{kin} / A + \Delta E_{ex} /A$ can be plotted as a function of $x'$ for given electron-electron coupling $g$ and doping $Q_d$ (see Fig.~\ref{fig4}).

\begin{figure}[tbh]
\includegraphics[width=.48\textwidth]{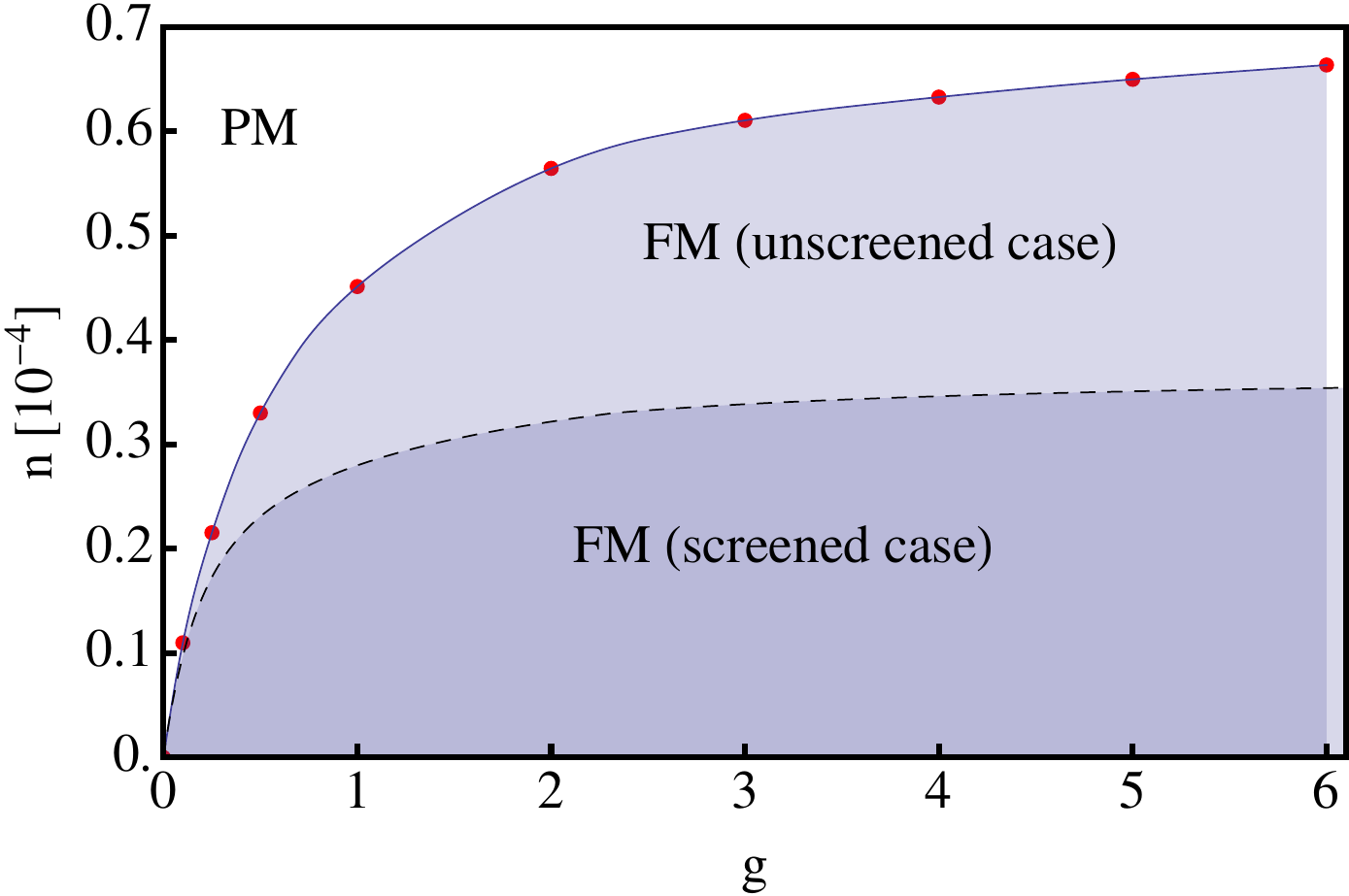}
\caption{(Color online) Phase diagram for $ABC$-trilayer graphene, in the case of an unscreened potential (solid blue line) and a screened potential (dashed black line). The red dots are the calculated values, while the solid line is an interpolation function based on the calculated points.}
\label{fig5}
\end{figure}

\noindent The minimum of $\Delta E(x)/A$ is estimated numerically by interpolation of points close to the minimum. The critical doping, where the minimum $\Delta E_{min}/A$ of $\Delta E(x)/A$ is zero, is found numerically by solving $\Delta E_{min}(Q_d)/A=0$. Since each minimum is a time consuming calculation, a simple binary search pattern is used (see inset of Fig.~\ref{fig4}).
\\\\

\subsection{Phase diagram}
\label{sec:phasediagram}

For a fixed value of $g=6$, we see in Fig.~\ref{fig4} the behavior of $\Delta E$ as a function of pocket sizes, upon varying the doping $Q_d$. For some doping values, $\Delta E$ is positive definite (paramagnetic phase), while for others $\Delta E$ attains a negative minimum (ferromagnetic phase). Inspection of the critical curve (thick line) shows that there is a first order phase transition between the paramagnetic and ferromagnetic phases. Repeating the entire procedure for different values of $g$ leads to the $g$ versus $n$ phase diagram depicted in Fig.~\ref{fig5}, where $n=Q_d^2/2$. The continuous solid line is an interpolation function of the calculated points.

These results were obtained by neglecting higher order corrections that lead to screening of the Coulomb potential. These effects will be considered in the next section.

\section{Effects of screening}
\label{sec:effectsscreening}

\subsection{Screened potential}

\begin{figure}[tbh]
\includegraphics[width=.48\textwidth]{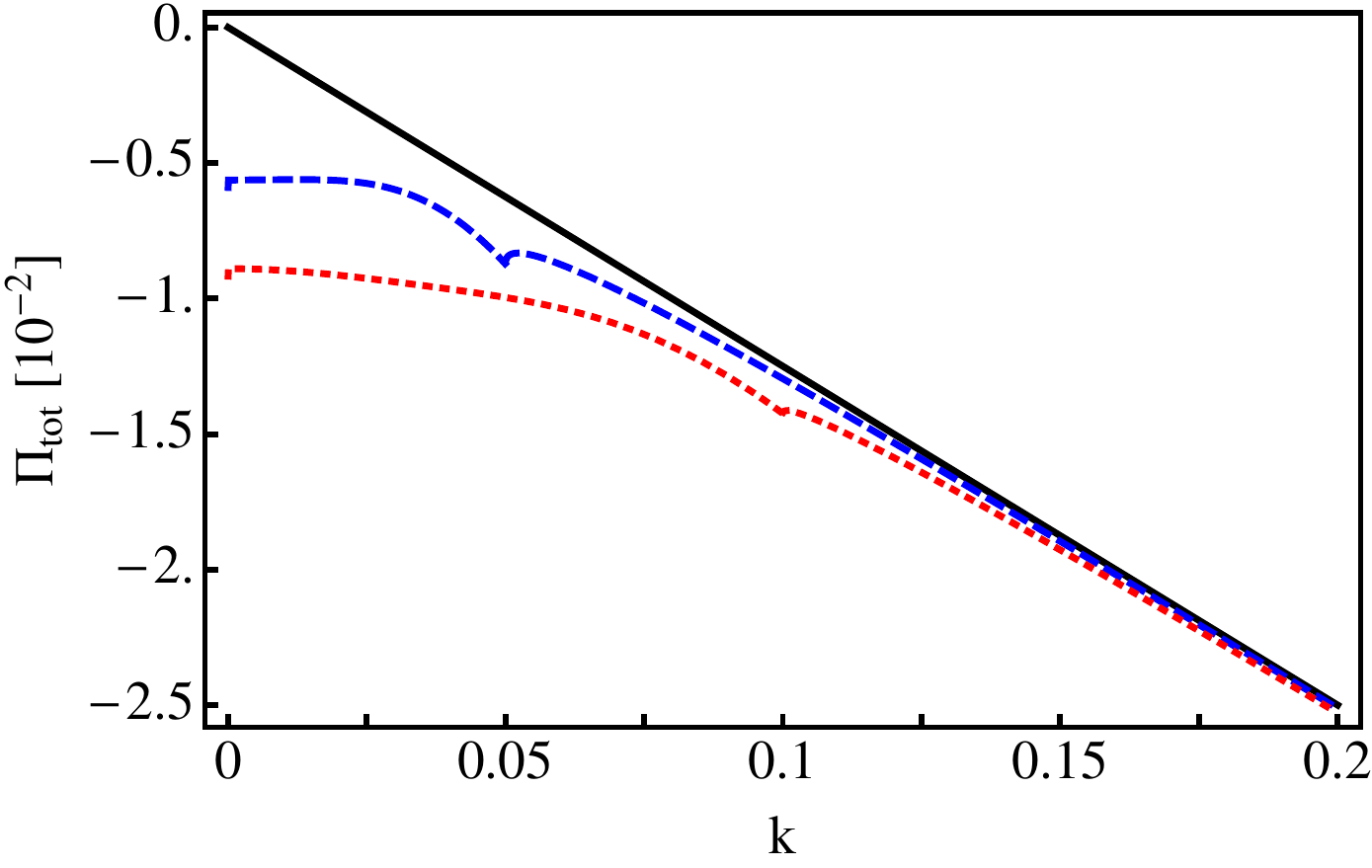}
\caption{(Color online) Plot of the bilayer graphene polarization $\Pi(k_F,\mathbf{k},0)$. The analytical expression derived in Ref.~\onlinecite{Gamayun2011} is shown as a blue dashed line for $k_F=0.025$ and as a red dotted line for $k_F=0.05$. The black solid line is the linear asymptote $\kappa k$ valid at large $k$, which we here extrapolate to small $k$.}
\label{fig6}
\end{figure}

Fourier transforming the real-space potentials $V^D$, $V^{ND}$ and $V^{2ND}$ and going to dimensionless variables yields
$$V^D=\frac{2 \pi g}{k}, \quad V^{ND}=\frac{2 \pi g e^{-kd}}{k}, \quad V^{2ND}=\frac{2 \pi g e^{-2kd}}{k},$$
where $g=e^2/\epsilon \hbar v_F$. As can be seen from Eq.~(\ref{potmatrix}) in the Appendix, the bare interaction line of $ABC$-trilayer graphene becomes a matrix $V_{mn}$, where $m$ and $n$ are layer indices. Therefore, the RPA renormalization of the potential\cite{Bruus2004} can be described by the Dyson-like equation
\begin{equation}
	\Diagram{\vertexlabel_m g \vertexlabel_n \\ \feynstrut{0.05}{0.05} \strut \\ g} = \Diagram{\vertexlabel_m g \vertexlabel_n} + \sum_{rl} \hspace{1mm}\Diagram{ \vertexlabel_m g \vertexlabel_r c \vertexlabel_l}\Diagram{g \vertexlabel_n \\ \feynstrut{0.05}{0.05} \strut \\ g}\text{  },
	\label{dysonlikeeq}
\end{equation}
where $r$ and $l$ are layer indices. Let $\mathcal{V}_{mn}$ be the renormalized potential. Then,
$$\sum_{rl} \hspace{1mm} \Diagram{ \vertexlabel_m g \vertexlabel_r c \vertexlabel_l}\Diagram{g \vertexlabel_n \\ \feynstrut{0.05}{0.05} \strut \\ g} = -\sum_{rl} V_{mr}(\mathbf{k}) \Pi_{rl}(\mathbf{k},i\omega) \mathcal{V}_{ln}(\mathbf{k},i\omega),$$
where
\begin{align*}
	\Pi_{rl}(\mathbf{k},i\omega) :&= \sum_\sigma \int \frac{d\mathbf{k}'}{(2\pi)^2} \int \frac{d\omega'}{\beta} G_0^{\sigma,rl}(\mathbf{k}',i\omega') \\
	&\times G_0^{\sigma,lr}(\mathbf{k}'-\mathbf{k},i\omega'-i\omega)
\end{align*}
and $G_0^{\sigma,lr}$ is the non-interacting Green's function of the system. Eq.~(\ref{dysonlikeeq}) is difficult to solve due to the layer dependence. However, for sufficiently low momenta $e^{-kd}\sim1$ and $e^{-2kd}\sim1$, which means that $V_{ij}\sim V\equiv 2 \pi g /k$. Thus, the layer dependence is removed, and Eq.~(\ref{dysonlikeeq}) can be solved with respect to $\mathcal{V}_{ij} \equiv  \mathcal{V}$:
\begin{equation}
	\mathcal{V}(\mathbf{k}) = \frac{V(k)}{1-V(k)\sum_{rl} \Pi_{rl}(\mathbf{k},i\omega)}.
	\label{renpotential}
\end{equation}

\subsection{Phase diagram}
Notice that Eq.~(\ref{renpotential}) does not converge to the true unscreened potential $V_{mn}$ as $\Pi_{rl}\rightarrow 0$. In order to achieve such a convergence, Eq.~(\ref{renpotential}) must be changed to
$$\mathcal{V}_{mn}(\mathbf{k}) = \frac{V_{mn}(k)}{1-V(k)\Pi^{tot}(\mathbf{k},i\omega)},$$
where
$$\Pi^{tot}(\mathbf{k},i\omega) \equiv  \sum_\sigma \Pi(Q_\sigma,\mathbf{k},i\omega) \equiv  \sum_{rl,\sigma} \Pi_{rl}(Q_\sigma,\mathbf{k},i\omega).$$
Since we are only interested in the long wavelength behavior, then $\omega \rightarrow 0$. For both, monolayer and bilayer graphene, the polarization $\Pi(Q_\sigma,\mathbf{k},0)$ behaves linearly in $k$ for large $k$, independent of Fermi momentum $Q_\sigma$, and exhibits an identical slope.\cite{Gamayun2011} This occurs because the dispersions are linear in the large-$k$ limit for both systems, and the Green's functions depend on the dispersion. Since the dispersion of $ABC$-trilayer graphene is also linear in the large-$k$ limit with the same slope as of the single- and bilayer dispersions, it is reasonable to assume that the linear behavior of $\Pi(Q_\sigma,\mathbf{k},0)$ is also present for $ABC$-trilayer graphene. In the exchange energy integrations, there are terms that are integrated from zero to the edge of the Brillouin zone (i.e. the cutoff $\Lambda=1$). Therefore, we will first focus on the screening effects coming from the linear behavior of $\Pi(Q_\sigma,\mathbf{k},0)$ and approximate it by $\Pi(Q_\sigma,\mathbf{k},0)=\kappa k$. An analytical expression of $\Pi(Q_\sigma,\mathbf{k},0)$ was calculated by Gamayun\cite{Gamayun2011} for bilayer graphene and is plotted in Fig.~\ref{fig6} for two values of the Fermi momentum (dashed and dotted lines) and compared with the linear estimate, where $\kappa \approx -0.12495$ (solid line).\cite{footnote1} Notice that the high-$k$ approximation that we use here is better than the one obtained using a two-band low-$k$ approximation. Indeed, for bilayer graphene where both the two-band and the full band polarizations were calculated, we see that the low-$k$ approximation of the two-band model misses the correct high-$k$ linear asymptotics and introduces a large error in the integrals which are performed up to the cutoff $\Lambda$.

Let us now use the linear expression for $\Pi$ and define $V\equiv g\tilde{V}$. Then, since $\Pi^{tot} V=2\Pi V$ is constant in $k$, the renormalized potential can be written as $\mathcal{V}_{ij} = \tilde{g} \tilde{V}_{ij}$, where
$$\tilde{g} \equiv  \frac{g}{1-4\pi g \kappa}.$$
Thus, the large momentum behavior of the renormalized potential effectively renormalizes the electron-electron coupling $g$. Let $n(g)$ be an interpolation function representing the phase boundary in the case of no screening (the solid line in Fig.~\ref{fig5}). Then,
$$n'(g) = n(\tilde{g}) = n\left( \frac{g}{1-4\pi g \kappa} \right),$$
is the phase boundary in the screened case. This boundary is shown by a dashed line in Fig.~\ref{fig5}. 

The low-$k$ regime of the polarization $\Pi(Q_\sigma,\mathbf{k},0)$ for $ABC$-trilayer graphene can be approximated by a constant $w=-1/[6 \pi k_F \beta]$, where $\beta = 400$.\cite{Rvg12} We will consider the case where $g=6$ and use the critical doping $k_F\approx 0.0116$ (see Fig.~\ref{fig4}), which leads to $w\approx -0.0114$. It is natural to let the transition into the linear regime of $\Pi$ occur at the point where $\kappa k_0=w$, i.e. at $k_0=0.1$ for $g=6$. The renormalized potential for $k<k_0$ now becomes
$$\mathcal{V}_{nm}(k)=\tilde{V}_{nm}(k) \frac{g}{1-4 \pi g w / k } \equiv \tilde{V}_{nm}(k)\tilde{\tilde{g}}(k).$$
As a crude approximation, we can let 
\begin{align*}
	\mathcal{V}_{nm} &\approx \tilde{V}_{nm} \text{ avg}_{k\in \Omega} \left[\tilde{\tilde{g}}(k)\right] \\
	&= \frac{\tilde{V}_{nm}}{\text{vol }\Omega} \int_0^{k_0} dp \int_0^{k_0} dp' \int_0^{2\pi} d\theta \frac{g}{1-4 \pi g w /k(p,p',\theta)},
\end{align*}
where $\Omega = [0,k_0]^2 \times [0,2\pi]$ is the domain where the constant regime of the polarization holds and $k \equiv \sqrt{p^2+p'^2-2pp'\cos \theta}$. Thus, the renormalized $g$ for $g=6$ at the critical $k_F$ becomes
$$\tilde{\tilde{g}} = \text{ avg}_{k\in \Omega} \left[\tilde{\tilde{g}}(k)\right] \approx 0.46,$$
where the integration was calculated numerically. At the same values of $g$ and $k_F$, the polarization in the linear regime yields $\tilde{g}\approx 0.58$. Thus, $\tilde{\tilde{g}}\sim \tilde{g}$, which implies that, as a first approximation, we may consider the linear approximation of the polarization for all momenta, which leads to the phase boundary represented by the black dashed line of Fig.~\ref{fig5}.

\section{Conclusions}
\label{sec:conclusions}

In this paper we study the magnetic properties of $ABC$-trilayer graphene using a tight-binding approach, where only the nearest-neighbor hopping parameters are taken into account. We include the Coulomb interaction and evaluate the exchange energy (Fock term) allowing for an unequal filling of the spin-up and spin-down bands. Then we calculate numerically the difference in energy between paramagnetic and ferromagnetic configurations of the system and identify the points of phase transitions for fixed values of the interaction parameter $g$. By repeating the calculations for several values of $g$, we obtain the phase diagram in the electron-electron coupling vs. doping plane. As a first step, we did not take into account the effects of Coulomb screening. The results are shown as the solid line in Fig.~\ref{fig5}.

Although the phase diagram for monolayer,\cite{PeCaNe05a} bilayer, \cite{NiCaNe06a} and $ABA$-trilayer\cite{Rvg11} graphene have been previously derived, effects of screening have been neglected until now. Our work represents the first step to incorporate these important effects.

For the unscreened case, at $g\approx 2.1$, a comparison with unscreened bilayer graphene\cite{NiCaNe06a} shows that $ABC$-trilayer graphene has a ferromagnetic behavior which is approximately 50 times stronger. Furthermore, a similar comparison with $ABA$-trilayer graphene\cite{Rvg11} shows that $ABC$-trilayer has a ferromagnetic behavior that is approximately 300 times stronger. At $g\approx 2.1$, monolayer graphene shows a paramagnetic behavior at all doping levels. In order for phase transitions to be present in monolayer graphene, the electron-electron coupling needs to exceed $g\approx 5$.\cite{PeCaNe05a} Fig.~\ref{fig4} shows that at $g=6$, the phase transition in $ABC$-trilayer graphene is of first order. This behavior persists for all couplings $g<6$. $ABA$-trilayer\cite{Rvg11} and bilayer graphene\cite{NiCaNe06a} also exhibits first order phase transitions for couplings $g<6$. This is in contrast to monolayer graphene, where both first order and second order phase transitions take place at given couplings $g$.\cite{PeCaNe05a} Thus, $ABC$-trilayer graphene behaves in a similar manner to bilayer and $ABA$-trilayer graphene, but exhibits a much stronger ferromagnetic behavior, making it easier to experimentally detect ferromagnetism. At $g=2.1$, the phase transition to ferromagnetism occurs at $n\approx 5.5\cdot 10^{-5}$. In SI-units the doping level becomes $\tilde{n}=g_s g_v \Lambda^2 Q_d^2/[4\pi] =g_sg_v Q_d^2/[2A]=n g_sg_v/A$, where $g_s=2$ and $g_v=2$ are the spin and valley degeneracies, respectively, and $A\approx 5.2 \cdot 10^{-16}$ $\text{cm}^2$ is the area of the Brillouin zone. Thus, neglecting valley degeneracy, $\tilde{n}\approx 2\cdot 10^{11}$ $\text{cm}^{-2}$. Note that, by mapping the parameter $x'$ of Fig.~\ref{fig4} to $x$, we see that the critical curve attains a minimum at $x<0$. Thus, in the ferromagnetic regime at $g=6$, the energy is always minimized for a configuration with two types of charge carriers. This behavior persists for all $g<6$.

These conclusions were reached by neglecting Coulomb screening. However, due to the diverging density of states in $ABC$-trilayer graphene, screening plays a very important role and must be taken into account. A thorough calculation of the polarization bubble in the full-band model is beyond the scope of this paper, and will be deferred to a future publication.\cite{Rvg12} Nevertheless, we have included screening effects within a simplified model. In the case of monolayer and bilayer graphene, the large-$k$ behavior of the bubble diagrams are linear in $k$, with the same slope $\kappa$. Arguing that this linear behavior also applies to $ABC$-trilayer graphene, and approximating the low-$k$ behavior of the polarization by a constant, we found that screening effects can be incorporated via a simple renormalization of the electron-electron coupling $g$. Fig.~\ref{fig5} shows that the large momentum behavior of the screening leads to a reduced ferromagnetic region in the $ABC$-trilayer graphene phase diagram. However, ferromagnetism is still approximately 25 times stronger than in unscreened bilayer graphene, which means that $ABC$-trilayer remains the material with the strongest ferromagnetic behavior.

We are aware that next-nearest neighbor hopping parameters, like $\gamma_3$ of the SWMc model can be of the same order as $\gamma_1$,\cite{FanZhang10} and that this can have an influence on the low-momentum behavior of the model. This parameter has been systematically neglected in studies of ferromagnetism in multilayer graphene (see Ref.~\onlinecite{NiCaNe06a} for bilayer and Ref.~\onlinecite{Rvg11} for $ABA$-trilayer). The reason is that, for studying the effects of other hopping parameters, one needs to redefine what is meant by a particle and a hole pocket due to the broken rotational symmetry of the dispersion around the K-point of the Brillouin zone, resulting from the SWMc model.\cite{FanZhang10,Rvg10} Furthermore, this broken symmetry leads to more complex integration boundaries, which makes the resulting numerical integrations intractable. 

Recently, an intrinsic bandgap of $6$ meV was experimentally observed in suspended $ABC$-trilayer graphene, and it was argued that it should be driven by interactions.\cite{Bao11} However, this gap did not appear in most of the samples placed on a substrate, which were investigated during the same study. Since suspended samples are more susceptible to ripples and deformations, it can well be that the spatial inversion symmetry was broken by strain, resulting in the intrinsic bandgap. Our studies should then apply for $ABC$-trilayer graphene on a substrate, without deformations. Because the dielectric constant is larger for samples on a substrate than for suspended samples (in vacuum), the coupling constant $g$ will be renormalized by a factor $\epsilon \sim 2.5$ for graphene on a SiO$_2$ wafer. Otherwise, the paramagnetic-ferromagnetic phase transition remains unaltered.

A simplified theoretical model which includes only on-site interactions suggests that the difference in bandstructure between $ABA$- and $ABC$-stacked trilayers should be enough to explain the presence of a gap due to antiferromagnetism in $ABC$ samples, while $ABA$-stacked trilayers remain ungapped.\cite{Xu12} These studies, however, cannot explain why the gap arises only in suspended samples.

Here we include long-range Coulomb interactions and investigate also the effect of screening. It is usually argued (without further ado) that screening is more important in $ABC$-trilayer than in the other related compunds. Our studies reveal that this is not always true, since the polarization is linearly increasing in a considerable region, over which one must integrate to obtain the exchange energy. This feature is similar in monolayer, bilayer, and $ABC$- trilayer graphene, and it is simply a consequence of the linear dispersion at intermediate values of $k$, which occurs in all the cases. Our studies reveal that the low energy approximation for the polarization is not always enough to ground fast conclusions. Although the final understanding about $ABC$-trilayer graphene has not yet been reached, we hope that our work will pave the way to possible extensions of the existing models for the investigation of ferromagnetism in multi-layer graphene using numerical methods.\\\\

\section*{Acknowledgments}
The authors acknowledge financial support from the Netherlands Organization for Scientific Research (NWO), as well as useful discussions with D. S. L. Abergel, D. Campbell, A. H. Castro Neto, and G. Japaridze.

\appendix

\section{Exchange energy}

The interaction Hamiltonian for $ABC$-trilayer graphene is shown in Eq.\ (\ref{intrealsp}). Fourier transforming $\rho_n$ and ${\Psi}_{\sigma,n}$ leads to
\begin{align}
	\nonumber \rho_n (\mathbf{q}) &= \int d\mathbf{r} \rho_n(\mathbf{r}) e^{-i\mathbf{q}\cdot\mathbf{r}} =  \sum_\sigma \int d\mathbf{r} {\Psi}^\dagger_{\sigma,n}(\mathbf{r}){\Psi}_{\sigma,n}(\mathbf{r}) e^{-i\mathbf{q}\cdot\mathbf{r}} \\
	\label{fourierrho} &= \frac{1}{A} \sum_{\mathbf{k},\sigma} {\Psi}^\dagger_{\sigma,n}(\mathbf{k} + \mathbf{q}){\Psi}_{\sigma,n}(\mathbf{k}), 
\end{align}
in the discrete limit. Using Eq.\ (\ref{fourierrho}) and Fourier transforming $V^D$, $V^{ND}$ and $V^{2ND}$ in Eq.\ (\ref{intrealsp}), going to the discrete limit, and subsequently rewriting the resulting expression into a matrix form yields
$$H_I = \frac{1}{2A} \sum_{\mathbf{q} \neq 0} \begin{pmatrix} \rho_1(-\mathbf{q}) & \rho_2(-\mathbf{q}) & \rho_3(-\mathbf{q}) \end{pmatrix} M \begin{pmatrix} \rho_1(\mathbf{q}) \\ \rho_2(\mathbf{q}) \\ \rho_3(\mathbf{q}) \end{pmatrix},$$
where (by omitting the $\mathbf{q}$ dependence for brevity)
\begin{align}
	M &= M_t+M_r \label{potmatrix} \\
	 &= \begin{pmatrix} V^D & V^{ND} & 0 \\ V^{ND} & V^D & V^{ND} \\  0 & V^{ND} & V^D \end{pmatrix} + \begin{pmatrix} 0 & 0 & V^{2ND} \\ 0 & 0 & 0 \\ V^{2ND} &  0 & 0 \end{pmatrix}. \nonumber
\end{align}
The matrix $M_t$ is diagonalized by $U_t$ such that $U_t^TD_tU_t = M_t$ where
$$U_t = \frac{1}{2} \begin{pmatrix} -\sqrt 2 & 0 & \sqrt 2 \\ 1 & -\sqrt 2 & 1 \\ 1 & \sqrt 2 & 1 \end{pmatrix}, \quad D_t = \begin{pmatrix} v_1 & 0 & 0 \\ 0 & v_2 & 0 \\ 0 & 0 & v_3 \end{pmatrix},$$
with $v_1\equiv V^D$, $v_2\equiv V^D-\sqrt 2 V^{ND}$ and $v_3\equiv V^D + \sqrt 2 V^{ND}$. Similarly, $M_r$ is diagonalized by $U_r$ such that $U_r^TD_rU_r = M_r$ where
$$U_r = \begin{pmatrix} 0 & 1 & 0 \\ -1/\sqrt 2 & 0 & 1/\sqrt 2 \\ 1/\sqrt 2 & 0 & 1/\sqrt 2 \end{pmatrix}, \text{  } D_r = \begin{pmatrix} 0 & 0 & 0 \\ 0 & -V^{2ND} & 0 \\ 0 & 0 & V^{2ND} \end{pmatrix}.$$
Let us define
\begin{align*}
	\frac{1}{\sqrt 2} \begin{pmatrix} \tilde{\rho}_1 \\ \tilde{\rho}_2 \\ \tilde{\rho}_3 \end{pmatrix} :&= U_t \begin{pmatrix} \rho_1 \\ \rho_2 \\ \rho_3 \end{pmatrix} = \frac{1}{\sqrt 2} \begin{pmatrix} -\rho_1 + \rho_3 \\ \rho_1/\sqrt 2 - \rho_2 + \rho_3/\sqrt 2 \\ \rho_1/\sqrt 2 + \rho_2 + \rho_3/\sqrt 2 \end{pmatrix}, \\
	\frac{1}{\sqrt 2} \begin{pmatrix} \tilde{\rho}_4 \\ \tilde{\rho}_5 \\ \tilde{\rho}_6 \end{pmatrix} :&= U_r \begin{pmatrix} \rho_1 \\ \rho_2 \\ \rho_3 \end{pmatrix} = \frac{1}{\sqrt 2} \begin{pmatrix} \sqrt 2 \rho_2 \\ -\rho_1 + \rho_3 \\ \rho_1 + \rho_3 \end{pmatrix}.
\end{align*}
Using the above diagonalizations yields
\begin{widetext}
\begin{align*}
	H_I &= \frac{1}{2A} \sum_{\mathbf{q} \neq 0} \frac{2\pi e^2}{\epsilon q} \frac{1}{\sqrt 2} \begin{pmatrix} \tilde{\rho}_1(-\mathbf{q}) & \tilde{\rho}_2(-\mathbf{q}) & \tilde{\rho}_3(-\mathbf{q}) \end{pmatrix} \begin{pmatrix} 1 & 0 & 0 \\ 0 & 1 - \sqrt 2 e^{-qd} & 0 \\ 0 & 0 & 1 + \sqrt 2 e^{-qd} \end{pmatrix} \frac{1}{\sqrt 2} \begin{pmatrix} \tilde{\rho}_1(\mathbf{q}) \\ \tilde{\rho}_2(\mathbf{q}) \\ \tilde{\rho}_3(\mathbf{q}) \end{pmatrix} \\
	&+\frac{1}{2A} \sum_{\mathbf{q} \neq 0} \frac{2\pi e^2}{\epsilon q} \frac{1}{\sqrt 2} \begin{pmatrix} \tilde{\rho}_4(-\mathbf{q}) & \tilde{\rho}_5(-\mathbf{q}) & \tilde{\rho}_6(-\mathbf{q}) \end{pmatrix} \begin{pmatrix} 0 & 0 & 0 \\ 0 & -e^{-2qd} & 0 \\ 0 & 0 & e^{-2qd} \end{pmatrix} \frac{1}{\sqrt 2} \begin{pmatrix} \tilde{\rho}_4(\mathbf{q}) \\ \tilde{\rho}_5(\mathbf{q}) \\ \tilde{\rho}_6(\mathbf{q}) \end{pmatrix}.
\end{align*}
\end{widetext}
By defining
$$ V_1(q) \equiv  \frac{\pi e^2}{\epsilon q}, \quad V_{2/3}(q) \equiv  \frac{\pi e^2}{\epsilon q}(1\mp \sqrt 2 e^{-qd}),$$
$$V_4(q) \equiv  0, \quad V_{5/6}(q) \equiv  \frac{\pi e^2}{\epsilon q}(\mp e^{-2qd}),$$
the Hamiltonian reduces to the compact form
$$H_I = \frac{1}{2A} \sum_{\mathbf{q} \neq 0} \sum_{s=1}^6 \tilde{\rho}_s (-\mathbf{q}) V_s (\mathbf{q}) \tilde{\rho}_s (\mathbf{q}).$$
Inspection of the operators $\tilde{\rho}_s$ for $s = 1,2,\dots,6$, indicates that they are all linear combinations of $\rho_n(\mathbf{q})$ for $n=1,2,3$. Thus, using Eq.\ (\ref{fourierrho}) one obtains
\begin{align*}
	\tilde{\rho}_s (\mathbf{q}) &= \sum_\mathbf{k} \tilde{\Psi}^\dagger(\mathbf{k}+\mathbf{q}) \tilde{\chi}_s \tilde{\Psi}(\mathbf{k}) \\
	&= \sum_\mathbf{k} {\Phi}^\dagger(\mathbf{k}+\mathbf{q}) \mathcal{M}^\dagger(\mathbf{k} + \mathbf{q})\tilde{\chi}_s \mathcal{M}(\mathbf{k}) {\Phi}(\mathbf{k}),
\end{align*}
where $\mathcal{M}(\mathbf{q})$ is the diagonalizing matrix of the $ABC$-trilayer Hamiltonian, $\tilde{\Psi} \equiv  (\tilde{\Psi}_1, \tilde{\Psi}_2, \tilde{\Psi}_3)$, with $\tilde{\Psi}_n(\mathbf{q})\equiv {\Psi}_n(\mathbf{q})/\sqrt{A}$ being a two component dimensionless annihilation operator working on layer $n$. The operator ${\Phi}^\dagger$ contains the band creation operators of the six bands. Thus, the six matrices $\tilde{\chi}^\alpha$ are defined as
\begin{align*}
	\tilde{\chi}_{1/6} &\equiv  \begin{pmatrix} \mp \mathbf{1}_2 & 0 & 0 \\ 0 & 0 & 0 \\ 0 & 0 & \mathbf{1}_2 \end{pmatrix}, \quad \tilde{\chi}_4 \equiv  \begin{pmatrix} 0 & 0 & 0 \\ 0 & \sqrt 2 \mathbf{1}_2 & 0 \\ 0 & 0 & 0 \end{pmatrix}, \\
	\tilde{\chi}_{2/3} &\equiv  \begin{pmatrix} \mathbf{1}_2/\sqrt 2 & 0 & 0 \\ 0 & \mp \mathbf{1}_2 & 0 \\ 0 & 0 & \mathbf{1}_2/\sqrt 2 \end{pmatrix}, \\
\end{align*}
where $\tilde{\chi}_5 \equiv  \tilde{\chi}_1$. By defining $\chi^s \equiv  \mathcal{M}^\dagger(\mathbf{k} + \mathbf{q})\tilde{\chi}_s \mathcal{M}(\mathbf{k})$ the $ABC$-trilayer interaction Hamiltonian can be written as
\begin{align*}
	H_I &= \frac{1}{2A} \sum_{\mathbf{q} \neq 0} \sum_{\mathbf{p}, \mathbf{p}'} \sum_{s=1}^6 \sum_{\alpha,\beta,\mu,\nu=1}^6 {\Phi}^\dagger_\alpha (\mathbf{p} - \mathbf{q}) \chi^s_{\alpha\beta} (\mathbf{p} - \mathbf{q}, \mathbf{p}) \\
	&\times {\Phi}_\beta (\mathbf{p}) V_s (\mathbf{q}) {\Phi}_\mu^\dagger(\mathbf{p}' + \mathbf{q}) \chi^s_{\mu\nu}(\mathbf{p}' + \mathbf{q}, \mathbf{p}') {\Phi}_\nu (\mathbf{p}').
\end{align*}
Let
$$\vert \mathbf{N} \rangle = \prod_{\mathbf{k},\nu,\sigma} \left[ {\Phi}^\dagger_{\sigma,\nu}(\mathbf{k}) \right]^{N_{\mathbf{k},\sigma,\nu}} \vert 0 \rangle$$
denote a Fock state of the system, where $N_{\mathbf{k},\sigma,\nu}\in\{0,1\}$ is the occupancy of electrons in the momentum state $\mathbf{k}$ of energy band $\nu$ with spin $\sigma$. Then, to first order, the energy of the system is described by
$$E_{I} = \langle \mathbf{N} \vert : H_I  : \vert \mathbf{N} \rangle,$$
where $:\hspace{1mm}:$ denotes normal ordering. Working out the expectation value of $H_I$ results in two distinct contributions. These are the Hartree (direct) and the Fock (exchange) contributions. Because of the Jellium background the Hartree contribution vanishes and only the Fock contribution remains. Thus,
\begin{align*}
	E_{ex} &= \frac{1}{2A} \sum_{\mathbf{p},\mathbf{p}'} \sum_{s=1}^6 \sum_{\alpha,\beta=1}^6 \sum_{\sigma,a} \mathcal{A}_{\sigma,\alpha,\beta,a}(\mathbf{p},\mathbf{p}') \\
	&\phantom{=}\times \chi^s_{\alpha\beta} (\mathbf{p}, \mathbf{p}') V_s (\mathbf{p}' -\mathbf{p}) \chi^s_{\beta\alpha}(\mathbf{p}', \mathbf{p}),
\end{align*}
where
\begin{align*}
	\mathcal{A}_{\sigma,\alpha,\beta,a}&=-\langle\mathbf{N} \vert {\Phi}^\dagger_{\sigma,\beta}(\mathbf{p}) {\Phi}_{\sigma,\beta}(\mathbf{p}) {\Phi}^\dagger_{\sigma,\alpha}(\mathbf{p}') {\Phi}_{\sigma,\alpha}(\mathbf{p}') \vert \mathbf{N} \rangle \\
	&= -n_{\sigma, \beta, a}(\mathbf{p}') n_{\sigma, \alpha, a}(\mathbf{p}),
\end{align*}
and $n_{\sigma, \alpha, a}(\mathbf{p}')$ are Fermi occupation functions, which in the $T\rightarrow 0$ limit become Heaviside step functions representing the pocket configurations shown in Fig.~\ref{fig3}. Going to the continuum limit reproduces the result shown in Eq.\ (\ref{exchenergy}). For further information on the numerical methods used to solve the exchange integral, see Ref.~\onlinecite{RO12}.


\begin{thebibliography}{30}
\bibitem{footnote2} We are using the letter $a$ to denote both, the lattice constant and the valley degree of freedom, as it is usually done in the literature.
\bibitem{footnote1} There is a factor of $1/(2\pi)$ difference from the article of Gamayun, due to differing conventions of the Fourier transform.
\end{thebibliography}
\end{document}